\documentclass[a4paper, amsmath,reprint,amssymb,aps,prd, groupedaddress, superscriptaddress]{revtex4-1}

\usepackage{graphicx}
\usepackage{dcolumn}
\usepackage{bm}
\usepackage{array}
\usepackage{hyperref}

\begin{document}


\title{Parametrizing theories of gravity on large and small scales in cosmology}

\author{Timothy Clifton}
\affiliation{School of Physics \& Astronomy, Queen Mary University of London, UK.\\}
\author{Viraj A. A. Sanghai}
\affiliation{School of Mathematics \& Statistics, Dalhousie University, Canada.\\}


\begin{abstract}
We present a link between parametrizations of alternative theories of gravity on large and small scales in cosmology. This relationship is established using theoretical consistency conditions only. We find that in both limits the ``slip'' and ``effective Newton's constant'' can be written in terms of a set of four functions of time, two of which are direct generalizations of the $\alpha$ and $\gamma$ parameters from post-Newtonian physics. This generalizes previous work that has constructed frameworks for testing gravity on small scales, and is to the best of our knowledge the first time that a link between parametrizations of gravity on such very different scales has been established. We expect our result to facilitate the imposition of observational constraints, by drastically reducing the number of functional degrees of freedom required to consistently test gravity on multiple scales in cosmology.
\end{abstract}


\maketitle

\textbf{\emph{Introduction} -- } There has recently been a lot of interest in testing the validity of Einstein's theory of General Relativity (GR) using cosmological observables \cite{review}. To fully exploit such tests requires us to understand the predictions that alternative theories of gravity could make for observations made over a wide range of scales, and to develop suitable frameworks for parametrizing these phenomena. This is a challenging prospect, as the physics involved in horizon-sized cosmological perturbations is quite different to that which occurs on smaller scales, where galaxies and clusters of galaxies are present.

We show that it is possible to relate the functions that parametrize gravity on non-linear scales ($k \gtrsim 10^{-2}\, {\rm Mpc}^{-1}$) to those that parametrize it on very large scales ($k\lesssim 10^{-4} \, {\rm Mpc}^{-1}$), where $k$ is the wavenumber of perturbations. Our approach to this problem is to lean heavily on the parametrized post-Newtonian (PPN) formalism \cite{PPN}, and to directly extend it to cosmology. Part of the reason for this is that the PPN approach has proven to be highly successfully in parametrising gravitational physics on solar system and astrophysical scales. Another is that the PPN formalism is known to encompass a wide array of alternative theories. Our final results give a parametrization of gravity in cosmology that, at leading order in perturbations, depends on only four functions of time. We call this approach parametrized post-Newtonian cosmology (PPNC) \cite{PPNC}.

We will consider perturbations about an FLRW background, which in longitudinal gauge has the scalar part
\begin{eqnarray}
\hspace{-2pt}
ds^2 = a^2 \Bigg[-(1-2\Phi) d\tau^2 + (1+ 2\Psi) \frac{(dx^2+dy^2+dz^2)}{\left( 1+\frac{1}{4}\kappa  r^2  \right)^2}\Bigg] \nonumber 
\end{eqnarray}
where $a=a(\tau)$ is the scale factor, $\tau=\int a^{-1} dt$ is conformal time, $\kappa$ is the spatial curvature parameter, and $r^2=x^2+y^2+z^2$. We then suppose that the linearized field equations that govern the metric potentials above can be written as
\begin{eqnarray}
&\hspace{-25pt}
\frac{1}{3}\nabla^2 \Psi - \mathcal{H}^2\Phi - \mathcal{H} \dot{\Psi}  + \kappa \Psi  = -\frac{4\pi G }{3} \, \mu \, \delta\! \rho \, a^2   \label{pert1}\\[3pt]
&\hspace{-15pt}
\frac{1}{3}{\nabla}^2 \Phi + 2\dot{\mathcal{H}}\Phi +  \mathcal{H}\dot{\Phi} + \ddot{\Psi} +\mathcal{H} \dot{\Psi}   = -\frac{4\pi G }{3} \, \mu (1-\zeta) \, \delta\! \rho \, a^2 
 \label{pert2}
\end{eqnarray}
where $\mu=\mu(\tau ,\mathbf{x})$ and $\zeta=\zeta(\tau ,\mathbf{x})$ are yet to be determined functions of time and scale, where $\delta \rho$ is the perturbation to the energy density, and where dots denote $d/d\tau$ and $\mathcal{H}=\dot{a}/a$. This follows the same logic as the PPN formalism, where the consequences of additional gravitational degrees of freedom are encoded in the parameters $\mu$ and $\zeta$, rather than being including as extra terms that have no counterparts in Einstein's theory.

If $\mu=1$ and $\zeta=0$ then Eqs. (\ref{pert1})-(\ref{pert2}) can be seen to reduce to those of GR. In the quasi-static limit they also reduce to well-known ``effective gravitational constant'' and ``slip'' parameters, defined via \cite{slip1, slip2} \\[-0.5cm]
\begin{equation}
\mu = - \frac{\nabla^2 \Psi}{4\pi G\,  \delta\! \rho \, a^2} \qquad {\rm and} \qquad \zeta = 1-\frac{{\nabla}^2 \Phi}{\nabla^2 \Psi} \, . \\[-0.25cm]
\nonumber
\end{equation}
Alternative theories of gravity generically result in $\mu \neq 1$ and $\zeta \neq 0$. It is therefore of direct physical interest to constrain these parameters with observations, to learn about how gravity behaves on cosmological scales.

The purpose of the current paper is to relate the small and large-scale behaviour of these parameters to each other, and to the background expansion. This will be achieved through the application of the following three consistency criteria: \\[-0.625cm]
\begin{itemize}
\setlength\itemsep{-0.5pt}
\item[(i)] On small scales, the gravitational field of non-linear structures should be describable using the parametrized post-Newtonian formalism.
\item[(ii)] The global rate of expansion of the Universe should be compatible with gravitational fields of the non-linear structures within it.
\item[(iii)] On very large scales, the evolution of perturbations in any theory should be consistent with the expansion rate of the Friedmann solutions of that theory. \\[-0.625cm]
\end{itemize}
We expect these criteria to be applicable to a wide range of metric theories of gravity, which are in a suitable sense ``close'' to GR, and that both admit perturbed FLRW solutions and fit into the PPN framework (see Refs. \cite{PPN, PPNC} for explicit examples). We will consider the implications of each of these three conditions in turn, before returning to show that together they give the values of $\mu$ and $\zeta$, on both large and small scales, in terms of a set of just four unspecified functions of time: $\{ \alpha, \gamma, \alpha_c, \gamma_c \}$. The first two of these are, in fact, exactly the same as the $\alpha$ and $\gamma$ parameters that occur in the PPN formalism (now allowed to vary over cosmic time), while the second two are intrinsic to cosmology. We will refer to the full set of four functions as the PPNC parameters. We set $c=1$ throughout this paper. Greek letters are used to denote spatial indices, and Latin letters to denote space-time indices.\\

\vspace{-6pt} \noindent
\textbf{\emph{Condition} (i): \emph{A viable weak field} -- } The industry standard for investigating and constraining theories of gravity in the slow motion and weak field regime is the PPN formalism \cite{PPN}. This formalism is built on the post-Newtonian expansion of gravitational and matter fields, and is expected to be valid in the presence of all non-linear structures down to the scale of neutron stars \cite{unreasonable}. All reasonable gravitational potentials are then postulated, and (constant) parameters are included before each of them in the metric, in order to create a general geometry that can be both constrained by observations and compared to the predictions of individual theories.

It seems reasonable to assume that the gravitational fields of galaxies and clusters of galaxies should also be described by post-Newtonian expansions, as long as these systems are small enough that the Hubble flow across them is still much smaller than the speed of light (i.e. that they are sub-horizon sized). If this is the case, then we can directly apply the PPN formalism to describe the gravitational fields within their vicinity. Such a description can be shown, by direct transformation, to be isometric to a region of perturbed FLRW space-time \cite{Tim1, vaas1}. 

In this case the coordinate transformations $\tilde{t} \rightarrow t +\frac{1}{2} a \mathcal{H} r^2$ and $\tilde{x}^{\mu} \rightarrow a \, x^{\mu} (1+\frac{1}{4} \mathcal{H}^2 r^2)$ allow the scalar gravitational potentials in the perturbed FLRW geometry to be written in terms of those in the isometric perturbed Minkowski space as \cite{Tim1}
\begin{eqnarray}
\Phi &=& \frac{1}{2}h_{00} - \frac{1}{2}  \dot{\mathcal{H}} r^2 \label{phi}
\\
\Psi &=& \frac{1}{6} \delta^{\mu \nu} h_{\mu \nu} + \frac{1}{4} \left(   \mathcal{H}^2 +\kappa \right) r^2 \, , \label{psi}
\end{eqnarray}
where the post-Newtonian metric is given by $g_{ab} =\eta_{ab} + h_{ab}$. If the post-Newtonian gravitational fields are then taken to be given by the form they take in the PPN formalism, we have \cite{PPN,PPNC}
\begin{eqnarray}
h_{00} = 2 \alpha U +\frac{1}{3} \alpha_c \, \tilde{r}^2 \quad {\rm and} \quad
\delta^{\mu \nu} h_{\mu \nu}=6 \gamma U + \gamma_c \, \tilde{r}^2 \, , \nonumber
\end{eqnarray}
where $U$ is the Newtonian gravitational potential that satisfies $\tilde{\nabla}^2 U = - 4 \pi G \rho$, where $\rho$ is the energy density, and where tildes denote coordinates in the Minkowski space. The two additional terms, proportional to $\alpha_c$ and $\gamma_c$ in these equations, are introduced in order to include the effects of $\Lambda$, and any of the other slowly-varying (or constant) additional fields that are often introduced in alternative theories. This gives a fully specified form for the cosmological scalar gravitational potentials on small scales, where non-linear structures are present, in terms of a set of four familiar quantities  (see \cite{PPNC} for further explanation). 

The functions $\alpha$, $\gamma$, $\alpha_c$ and $\gamma_c$ that are used here should in general be expected to be functions of time only, if inhomogeneous gravitational fields are expected to be sourced by matter fields (see Ref. \cite{PPNC} for details). The reader will note that the PPN parameter $\alpha$ has been made explicit in the expression for $h_{00}$. This is usually set to unity when the PPN formalism is presented, so that the standard Newton-Poisson equation is recovered in the appropriate limit. Here we have made this parameter explicit as we only require the Newton-Poisson equation with the experimentally recovered value of $G$ to be applicable at the present moment of time. Earlier (or later) cosmological epochs may have different values of Newton's constant in alternative theories of gravity \cite{uzan}. The second PPN parameter, $\gamma$, affects Shapiro time delay and deflection of light rays, among other things, and is currently most strongly constrained from observations of radio signals from the Cassini spacecraft to be $\gamma=1+(2.1\pm2.3) \times 10^{-5}$ \cite{cassini}. For the case of GR, $\alpha = \gamma =1$ and $\alpha_c= -2 \gamma_c=\Lambda$ for all time.\\

\vspace{-6pt} \noindent
\textbf{\emph{Condition} (ii): \emph{A compatible Hubble rate} -- } The Hubble rate of the Universe's expansion is not independent of the gravitational fields of the objects within it. On the contrary, one can build explicit constructions in which the global expansion emerges from the Newtonian-level gravitational fields of the structures it contains \cite{Tim1}, with post-Newtonian fields giving small corrections \cite{vaas1,vaas2,thesis}. Such models make clear that the large-scale expansion of a cosmological model should be able to be parametrized with the same set of functions as weak gravitational fields on small scales. This fact was used in \cite{PPNC} to extract the large-scale expansion of a Universe that is governed by the PPN metric on small scales. Here we generalize this construction to universes constructed from regions of space with periodic boundary conditions, rather than the more restrictive reflection symmetry that was previously assumed.

First we take two spatial derivatives of Eqs. (\ref{phi}) and (\ref{psi}), and average the result over a large volume of space, to find
\begin{eqnarray}
\label{con}
\mathcal{H}^2 &=&  \frac{8\pi G}{3} \, \gamma \, \langle \rho \rangle \, a^2- \frac{2 \gamma_{c} \, a^2 }{3}- \kappa  \label{fried} \\[-2pt]
\dot{\mathcal{H}} &=& -\frac{4\pi G }{3} \, \alpha \, \langle \rho \rangle \, a^2 + \frac{\alpha_{c} \, a^2}{3} \, , \label{acc}
\end{eqnarray}
where angular brackets denote the spatial average of the quantity within, such that $\langle \cdots \rangle = \int \cdots d^3x/\int d^3x$. In deriving these equations we have used the result $\langle \nabla^2 \Phi\rangle = 0=\langle \nabla^2 \Psi\rangle $ for a region of space with periodic boundary conditions, and we have assumed that the PPNC parameters $\{\alpha, \gamma, \alpha_c, \gamma_c\}$ are not functions of space. Failure of these results to hold would indicate the presence of large cosmological back-reaction \cite{br}, and would invalidate the use of a global FLRW background, which is not the scenario that we wish to consider here.

Subtracting the averaged Eqs. (\ref{fried})-(\ref{acc}) from the original Eqs. (\ref{phi})-(\ref{psi}) then gives
\begin{eqnarray}
\hspace{-10pt}
\label{np1}
\nabla^2 \Phi = -4\pi G \, \alpha \, \delta\!\rho  \, a^2 \label{np1} \quad {\rm and} \quad
\nabla^2 \Psi = -4\pi G \, \gamma \, \delta\!\rho \, a^2 , 
\end{eqnarray} 
where $\delta\!\rho = \rho - \langle \rho \rangle$. It can be seen that the PPN parameters $\alpha$ and $\gamma$ parametrize both the matter terms in the Friedmann Eqs. (\ref{fried})-(\ref{acc}) and the small-scale cosmological perturbations in Eqs. (\ref{np1}). On the other hand, the cosmological parameters $\alpha_c$ and $\gamma_c$ occur only in the Friedmann equations, and take the place of the dark energy terms. 

The reader may note that in addition to the equations above, we have an additional integrability condition on $\{\alpha, \gamma, \alpha_c, \gamma_c \}$ given by
\begin{equation}
4\pi G\langle {\rho} \rangle  = \left(\alpha_{c} + 2\gamma_{c} + \gamma^{\prime}_{c} \right)\big/ \left(\alpha - \gamma + \gamma^{\prime} \right)\, , 
\label{addcon1}
\end{equation} 
where ${}^{\prime} = d/d \ln a$. The derivation of this equation uses the result $\langle \rho \rangle \propto a^{-3}$, which can be obtained from averaging the energy-momentum conservation equations, as well as the standard assumption that we do not have any interactions between matter fields at the level of the background cosmology. Explicit expressions for these PPNC parameters, for scalar-tensor and vector-tensor theories of gravity, were derived in \cite{PPNC}. The following sections extend the domain of validity of this work to horizon-sized scales, and so give information on the scale dependence of modified gravity parameters.\\

\vspace{-6pt} \noindent
\textbf{\emph{Condition} (iii): \emph{Large-scale perturbations} -- } On sufficiently large scales the spatial gradients of perturbations are expected to be become much smaller than their time derivatives. Neglecting their contribution to the field equations then results in a situation where only the time dependence of the gravitational potentials is of significance for their evolution. It therefore becomes possible to model these perturbations as Friedmann solutions with perturbed energy density and spatial curvature, as well as space and time coordinates. As noted by Bertschinger; the evolution of such perturbations must then be specified entirely by the Friedmann equations, and can be deduced without knowing any further information about the field equations of gravity \cite{Bertschinger:2006aw}.

In the present situation, this result is especially useful as our parametrized Friedmann Eqs. (\ref{fried})-(\ref{acc}) can now be used to determine the form of very-large-scale perturbations. The result must then also be dependent only on the PPNC parameters $\{\alpha,\gamma,\alpha_c,\gamma_c\}$. The first step in doing this is to perturb the FLRW background in spherical polar coordinates such that the conformal time $\tau \to \tau + \delta \tau$ and radial coordinate $\chi \to \chi(1+\delta_{\chi})$, and to perturb spatial curvature and energy density such that $\kappa \to \kappa(1+\delta_{\kappa})$ and $\langle \rho \rangle \to \langle \rho \rangle (1+ \delta)$, where $\delta\tau$, $\delta_{\chi}$, $\delta_{\kappa}$ and $\delta$ are all small quantities. Furthermore, the quantity $\delta_{\kappa}$ must also be a constant, as it is a perturbation of the (constant) spatial curvature. The $\chi$ coordinate here is such that the conformal part of the spatial line-element can be written as $d\bar{s}_{(3)}^2 = d\chi^2 + \hat{r}^2(\chi, \kappa) d\Omega^2$, where $d\Omega$ is the solid angle.

Writing the large-scale perturbation in terms of the transformed Friedmann solution then means that the following must be true \cite{Bertschinger:2006aw}:
\begin{equation}
\delta_{\chi} = -\frac{1}{2}\delta_{\kappa} \label{kappa}
\end{equation}
\begin{equation}
\Phi + \Psi = \frac{\partial}{\partial \tau}\bigg( \frac{\delta_{\chi}}{\mathcal{H}}\bigg(1- 2\frac{\partial \ln a}{\partial \ln \kappa}\bigg)  - \frac{\Psi}{\mathcal{H}}\bigg) + \delta_{\chi} \, , \label{condition5}
\end{equation}
where we have eliminated the dependence on $\delta \tau$, and where Eq. \eqref{kappa} tells us $\delta_{\chi}$ must be a constant. Using $\langle \rho \rangle \propto a^{-3}$, we can then Taylor expand $\delta$ to obtain $\delta= -3(\Psi - \delta_{\chi})$, and hence $\dot{\delta} = -3\dot{\Psi}$  (as expected from perturbing the continuity equation on very large scales). Using this result to replace $\delta_{\chi}$ in Eq. \eqref{condition5}, and performing further manipulations, one finds
\begin{align}
-\mathcal{H}^2\Phi - \mathcal{H} \dot{\Psi} +  {\kappa} \Psi
&=- \frac{\delta}{3} \left({\mathcal{H}^2 - \dot{\mathcal{H}}} + {\kappa}\right)  \label{lwmain1}
\\
\hspace{-0.1cm}  2\dot{\mathcal{H}} \Phi + \mathcal{H} \dot{\Phi} + \ddot{\Psi} +\mathcal{H} \dot{\Psi} 
&= \frac{\delta}{3} \left(\frac{2\dot{\mathcal{H}}\mathcal{H} - \ddot{\mathcal{H}} }{\mathcal{H}} \right) \, . \label{lwmain2}
\end{align}
Finally, we can simplify the right-hand sides of these equations by using the parametrized Friedmann Eqs. (\ref{fried})-(\ref{acc}), and the constraint in Eq. (\ref{addcon1}), to get
\begin{align}
&-\mathcal{H}^2\Phi - \mathcal{H} \dot{\Psi} +  {\kappa} \Psi \nonumber\\
=& -\frac{4\pi G }{3} \delta \rho \, a^2 \bigg(\gamma  - \frac{1}{3} {\gamma^{\prime}}   + \frac{1}{12\pi G  \langle \rho \rangle} {\gamma^{\prime}_{c}} \bigg)  \label{finalmain2}
\end{align}
and
\begin{align}
&2\dot{\mathcal{H}} \Phi + \mathcal{H} \dot{\Phi} + \ddot{\Psi} +\mathcal{H} \dot{\Psi} \nonumber \\
 =& -\frac{4\pi G }{3} \delta \rho \, a^2 \bigg(\alpha  - \frac{1}{3} {\alpha^{\prime}}   + \frac{1}{12\pi G  \langle \rho \rangle} {\alpha^{\prime}}_{c} \bigg) \, . \label{finalmain1}
\end{align}
These results can be compared to Eqs. (\ref{pert1})-(\ref{pert2}). They constitute a parametrization of super-horizon-sized fluctuations in terms of the PPNC parameters and their time derivatives only. The reader may note that nowhere in this derivation have we assumed anything about the field equations of gravity, except that they should result in the parametrized Friedmann Eqs. (\ref{fried})-(\ref{acc}), which themselves are a direct result of treating gravity on small scales as being governed by the PPN metric.\\

\vspace{-6pt} \noindent
\textbf{\emph{Results}}
-- We can now bring together the results above to demonstrate the link between parametrizations of gravity on large and small scales in cosmology. Comparing Eqs. (\ref{np1}) with (\ref{pert1})-(\ref{pert2}) allows the effective gravitational constant parameter, $\mu$, and the gravitational slip parameter, $\zeta$, to be related to the $\alpha$ and $\gamma$ parameters via \cite{PPNC}
\begin{eqnarray}
\lim_{k\rightarrow \infty} \mu  &=&\gamma \label{end1} \\
\lim_{k\rightarrow \infty} \zeta &=& 1 - \frac{\alpha}{\gamma} \label{end2} \, ,
\end{eqnarray}
where $k \rightarrow \infty$ corresponds to the small-scale limit. We expect this parametrization to be valid for $k \gtrsim 0.01\, {\rm Mpc}^{-1}$, where non-linear structures exist, and post-Newtonian expansions are expected to be valid.

\begin{table}[t!]
\begin{center}
{\renewcommand{\arraystretch}{1.25}
\begin{tabular}{|c|c||c|c|}
\hline
\, PPNC \, & \, observational \, & \, PPNC \, & \, observational \, \\[-5pt]
parameter & \, constraint \, & derivative & \, constraint \, \\
\hline
$\alpha$ & $1$ & $\alpha^{\prime}$ & $0 \pm 0.01$ \cite{uzan} \\
$\gamma$ & $1 \pm 10^{-5}$ \cite{cassini} & $\gamma^{\prime}$ & $0.0 \pm 0.1$ \cite{jou} \\
$\alpha_c / H_0^2$ & $2.07 \pm 0.03$ \cite{planck9} & $\alpha_c^{\prime} / H_0^2$ & $0.12 \pm 0.25$ \cite{planck9} \\
$\gamma_c / H_0^2$ & $-1.04 \pm 0.02$ \cite{planck9} & $\gamma_c^{\prime} / H_0^2$ & $-0.06 \pm 0.12$ \cite{planck9} \\
\hline
\end{tabular}
}
\caption{Observational constraints on the present-day values of the PPNC parameters $\{\alpha, \gamma, \alpha_{c}, \gamma_{c}\}$, and their derivative with respect to $\ln a$. Here we have taken $\alpha=1$ (by definition), and have assumed the $\Lambda$s that appear in both Friedmann equations are identical, with $\Omega_{\Lambda}= 0.69 \pm 0.01$ \cite{planck9}. We have used $H_0=68 \pm 1 \, {\rm km} \, {\rm s}^{-1} \, {\rm Mpc}^{-1}$ \cite{planck9} to find the relevant derivatives, and derived the constraints on $\alpha_c^{\prime}$ and $\gamma_c^{\prime}$ using the result $\omega_{\Lambda}+1=-0.02\pm 0.04$ \cite{planck9}. The $\alpha^{\prime}$ and $\gamma^{\prime}$ results come from adapting constraints on $\dot{G}/G$ and $\Sigma$ from Refs. \cite{uzan} and \cite{jou}. A more detailed study of observational constraints on these parameters will be presented in Ref. \cite{bs}.} \vspace{-10pt}
\label{tab1}
\end{center}
\end{table}

Similarly, comparing Eqs. (\ref{finalmain2})-(\ref{finalmain1}) with (\ref{pert1})-(\ref{pert2}) allows us to read off
\begin{eqnarray}
\lim_{k\rightarrow 0} \mu &=&\gamma  - \frac{1}{3} {\gamma}^{\prime}   + \frac{1}{12\pi G  \langle {\rho}\rangle} {\gamma_{c}^{\prime}} 
\label{end3} \\
\lim_{k\rightarrow 0} \zeta &=& 1-\frac{\alpha  -  {\alpha}^{\prime}/3   + {\alpha}_{c}^{\prime}/{12\pi G  \langle \rho \rangle}  }{\gamma  -  {\gamma}^{\prime}/3   +{\gamma^{\prime}_{c}}/ {12\pi G \langle \rho \rangle} } \ , \label{end4}
\end{eqnarray}
where primes again represent $d/d\, \ln a$, and where the spatial gradient terms in Eqs. (\ref{pert1})-(\ref{pert2}) are expected to be negligible. Here the limit $k \rightarrow 0$ indicates that this parametrization should be expected to be valid on super-horizon scales, such that $k \lesssim 0.0001\, {\rm Mpc}^{-1}$. The reader may note that the terms involving $ \langle \rho \rangle$ can be replaced by our parameters using Eq. (\ref{addcon1}), if required. 

\begin{figure}[b!]
\includegraphics[width=1.05 \columnwidth]{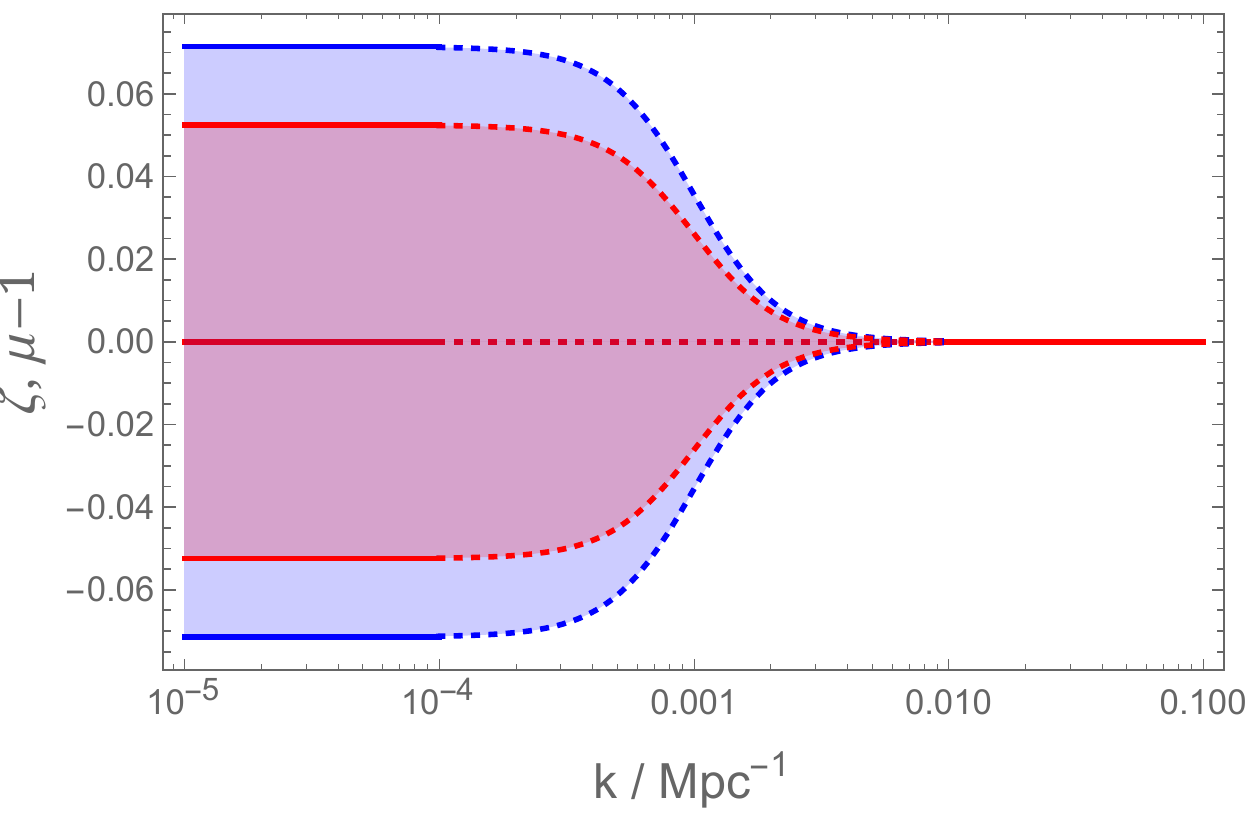}
\caption{The small-scale ($k \gtrsim 0.01\, {\rm Mpc}^{-1}$) and large-scale ($k\lesssim 10^{-4}\, {\rm Mpc}^{-1}$) limits of the $\zeta$ (red) and $\mu$ (blue) parameters at the present time, connected by an interpolating $\tanh$ function (dotted). We have used the values of $\{\alpha, \gamma, \alpha_{c}, \gamma_{c}\}$ and their associated errors and derivatives from Table \ref{tab1}. The shaded areas show the $1 \sigma$ confidence regions, where errors have been assumed to be independent.}
\label{fig1}
\end{figure}

The first thing that can be noted about Eqs. (\ref{end1})-(\ref{end4}) is that the large and small-scale parametrizations can both be written exclusively in terms of the PPNC parameters, $\{\alpha, \gamma, \alpha_{c}, \gamma_{c}\}$. These parameters also occur in the background Friedmann Eqs. (\ref{fried})-(\ref{acc}), and therefore provide a complete set (in addition to the usual cosmological parameters) that can parametrize gravitational physics over a wide range of scales. Further, the reader may note that if $\{\alpha, \gamma, \alpha_{c}, \gamma_{c}\}$ are all independent of time then the large and small-scale limits of the slip and effective Newton's constant are identical. This is not a surprise, as independence from spatial scale should be expected to lead to independence from time in metric theories of gravity. These results reduce to those expected from GR, when $\alpha = \gamma =1$ and $\alpha_{c} = -2 \gamma_c=\Lambda$, as expected \cite{Sophia0, Sophia1}.

Eqs. (\ref{end1})-(\ref{end4}) represent the main results of this paper. They extend parametrized gravity from small non-linear scales, to large horizon-sized scales. In order to show the power of this approach, we have considered indicative observational constraints on the individual PPNC parameters in Table \ref{tab1}. These are then subsequently used in Fig. \ref{fig1} to show how current bounds on the values of $\{\alpha, \gamma, \alpha_{c}, \gamma_{c}\}$ can be used to derive bounds on $\zeta$ and $\mu$ on both small {\it and} large scales. It is the possible time-variation of the PPNC parameters that is the reason for the difference between the value of $\zeta$ and $\mu-1$ over this range of $k$. The constant value of $\zeta$ and $\mu$ in the regions $k\lesssim 10^{-4} \, {\rm Mpc}^{-1}$ and $k \gtrsim 10^{-2}\, {\rm Mpc}^{-1}$ is due to the scale-independence of $\{\alpha,\gamma,\alpha_c,\gamma_c\}$, as discussed above. Due to the different values of $\zeta$ and $\mu$ in these two regions, we require scale dependence in the region $10^{-4} \, {\rm Mpc}^{-1}\lesssim k \lesssim 10^{-2}\, {\rm Mpc}^{-1}$. We have achieved this with an example interpolating curve in Fig. \ref{fig1}.\\

\vspace{-6pt} \noindent
\textbf{\emph{Conclusions} -- } We have obtained a direct link between horizon-sized cosmological perturbations, and those on smaller non-linear scales, in terms of a set of just four functions of time: $\{\alpha, \gamma, \alpha_{c}, \gamma_{c}\}$. This set of functions parametrize a wide array of minimal modifications to GR, of the type that has been long used to test gravity in the Solar System and binary pulsars. Here we have found that they can be used to describe gravitational physics in the Solar System, through astrophysical and small cosmological scales, all the way up to the super-horizon scales involved in considering the entire observable universe.

To the best of our knowledge, this is the first time that theories of gravity have been parametrized consistently on such a large range of scales, using such a compact set of parameters. These results can be contrasted with the parametrized post-Friedmannian (PPF) \cite{Hu1, Hu2, amin, PPF1, PPF2, PPF3} and effective field theory (EFT) \cite{eff1, eff2, eff3, eff4, eff5, eff6, eff7, eff8, eff9, eff10} approaches, which can contain larger numbers of unknown functions, and that often require the degrees of freedom in the theory to be specified from the outset. We expect our parametrization to be useful for testing minimal deviations from GR with future large-scale surveys \cite{ska, lsst, euclid}, and in particular for future precision constraints on gravity in cosmology. Future work will consider the effect of screening mechanisms and fifth forces in this approach \cite{Fleury}.\\

\vspace{-0.1cm}
\noindent
\textbf{{Acknowledgements:} } We are grateful to P. Bull and A. Pourtsidou for helpful discussions. VAAS acknowledges the support of Atlantic Association for Research in Mathematical Sciences, and TC of the STFC.

\appendix

\end{document}